\documentclass[a4paper,11pt]{article}
\usepackage{amssymb,amsmath}
\usepackage{upgreek}
\usepackage{secdot}
\usepackage{tocloft}
\usepackage{makecell}
\usepackage{chngcntr}
\counterwithout{equation}{section}

\pdfoutput=1 

\usepackage{jinstpub} 

\title{\boldmath Characterization of plastic scintillator bars using fast neutrons from $D$-$D$ and $D$-$T$ reactions.}

\author[a,b,1]{R.~Dey,}
\author[a]{P.~K.~Netrakanti,}
\author[a]{D.~K.~Mishra,}
\author[a]{S.~P.~Behera,}
\author[a,b]{D.~Mulmule,}
\author[c]{T.~Patel,}
\author[c]{\\P.~S.~Sarkar,}
\author[a,b]{V.~Jha,}
\author[a,b]{L.~M.~Pant}

\affiliation[a]{Nuclear Physics Division, Bhabha Atomic Research Centre,\\ Trombay, Mumbai, India - 400085.}
\affiliation[b]{Homi Bhabha National Institute, \\Anushakti Nagar, Mumbai, India - 400094.}
\affiliation[c]{Neutron \& X-ray Physics Division, Bhabha Atomic Research Centre,\\ Trombay, Mumbai, India - 400085.}

\emailAdd{ronidey@barc.gov.in}

\abstract{
We report results of fast neutron response in plastic scintillator (PS) bars from deuterium-deuterium ($D$-$D$) and deuterium-tritium ($D$-$T$) reactions using Purnima Neutron Generator Facility, BARC, Mumbai. These measurements are useful in context of Indian Scintillator Matrix for Reactor Anti-Neutrino (ISMRAN) detection, an array of 10$\times$10 PS bars, used to measure reactor anti-neutrinos through inverse beta decay (IBD) signal. ISMRAN detector, an above-ground experiment close to the reactor core ($\sim$13m), deals with an active fast neutron background inside the reactor hall. A good understanding of fast neutron response in PS bars is an essential pre-requisite for suppression and discrimination of fast neutron background from IBD events. A monoenergetic neutron beam from the fusion reaction of $D$-$D$ at 2.45 MeV and $D$-$T$ at 14.1 MeV are used to characterize the energy response in these bars. The neutron energy response function has been simulated using the GEANT4 package and are compared with the measured data. A reasonable agreement of deposited energies by fast neutrons in PS bars between data and simulation are obtained for these reactions. The ratio of energy deposition in adjacent bars is used to discriminate between prompt IBD, fast neutron and neutron capture cascade gamma events.
}

\keywords{Anti-neutrino, Plastic scintillator, Neutron generator, Neutron response.}
\begin{document}
\maketitle
\flushbottom
\section{{Introduction}:}
\label{sec:intro}
The experimental measurement of reactor-based electron type anti-neutrinos ($\overline\nuup_{e}$) has provided the key aspects in determining the mixing angles and mass splittings in the three flavour framework. Experimental collaborations such as Daya Bay~\cite{DBMIX}, RENO~\cite{RENO} and Double Chooz~\cite{DChooz} have reported the most precise  measurements of the mixing angle ($\mathrm{\theta_{13}}$) from reactor-based $\overline\nuup_{e}$ experiments. The ongoing effort for the search of sterile neutrinos at very short baseline experiments~\cite{DANSS,NEOS,RENOST,STEREO,PROSPECT,SOLID} have reported significant exclusions in the allowed parameter space of  $ \mathrm{sin^2(2\theta_{14})}$ and $\mathrm{\Delta m^2_{41}}$. Also measurements of reactor-based $\overline\nuup_{e}$ induced prompt positron energy spectrum shows a statistically significant excess of events over the prediction, particularly in the energy range of $\mathrm{5-7}$ MeV~\cite{DB5MeV,RENO5MeV,Huber5MeV}.

Indian Scintillator Matrix for Reactor Anti-Neutrino (ISMRAN) is an above-ground very short baseline reactor $\overline\nuup_{e}$ experiment at the Dhruva Reactor~\cite{DHRUVA} facility in Bhabha Atomic Research Centre (BARC)~\cite{ISMRANNIM}. It is designed to measure the yield and energy spectrum of $\overline\nuup_{e}$, via inverse beta decay (IBD) reaction, for monitoring fuel evolution ~\cite{IAEA,PANDA} and search for the existence of sterile neutrino with a mass on the order of $\sim$1 eV /$\mathrm{c^{2}}$~\cite{Shiba}. The excess of $\overline\nuup_{e}$ events in data from prediction at $\mathrm{5-7}$ MeV in the spectral shape of the measured positron energy spectrum will also be addressed using ISMRAN experiment. ISMRAN detector setup comprises of 100 plastic scintillator (PS) bars arranged in 10$\times$10 segmented geometry. 
Each PS bar has a dimension of $\mathrm{100 cm \times 10 cm \times 10 cm}$ and is wrapped with Gadolinium oxide($\mathrm{Gd_{2}O_{3}}$) coated on aluminized mylar foils. The areal density of the Gd on these foils is 4.8$~\mathrm{mg/cm^2}$. For the readout of scintillation photon signal, PMT with 3$''$ diameter is directly coupled to each end of the PS bar. 
The ISMRAN detector will be located at a distance of $\sim$13 m from the core of the Dhruva reactor. To minimize the influence of ambient backgrounds resulting from cosmic rays, natural radioactivity and reactor-related fast neutrons and photons ($\gamma$), the ISMRAN detectors are enclosed by a shielding made of 10 cm thick lead (Pb) and 10 cm thick borated  polyethylene (BP). From simulations it is estimated that a moderate shielding of 10 cm of Pb and 10 cm of BP has an acceptance of $\sim$10$\%$ for fast neutron events, within the energy range of 2 MeV to 10 MeV, inside the shielding~\cite{ISMRANNIM}. This choice of shielding is optimized by keeping the tonnage requirement of the massive shielding structures on the floor of the reactor hall.   
A muon veto system made of plastic scintillator paddles on the exterior of the shielding structure is used to veto the cosmic muon background. The complete setup will be housed on a movable trolley for allowing the movement of the ISMRAN detector to various distances from the reactor core. Further details about the schematics of the ISMRAN setup can be found in reference~\cite{ISMRANNIM}.
Regardless of the Pb and BP shielding, there will be some residual fast neutrons that can enter ISMRAN detector and strike PS bars to produce proton recoil events which can mimic the signature of prompt IBD signal which consists of positron energy loss through ionization and annihilation with an electron yielding two $\gamma$-rays inside the scintillator volume. Also these fast neutrons can get thermalized after generating a fake prompt signal and may get captured on Gd foil or hydrogen, which gives a delayed IBD signal. These fast neutron events would lead to an irreducible background, difficult to be discriminated from the real IBD event signatures~\cite{Reactor bkg}. Most of the reactor  $\overline\nuup_{e}$ experiments consist of liquid scintillators as the target detector. These detectors have better energy resolutions and can discriminate the fast neutron background using pulse shape discrimination technique as opposed to a segmented plastic scintillator detector like ISMRAN. A common technique used for identifying signal and background events by various experiments is to separate events based on the timing information between the prompt and delayed events~\cite{DBBkgd,Mei,Wulandari,bowden}. The PROSPECT experiment uses a highly segmented liquid scintillator setup for the detection of reactor $\overline\nuup_{e}$ with an excellent signal to background ratio~\cite{PROSPECT}. For a plastic scintillator based matrix for reactor $\overline\nuup_{e}$ detection, PANDA experiment uses the energy depositions in the segmented geometry to distinguish between prompt and delayed IBD events~\cite{PANDA}. A detailed discussion of measured signal to background ratio by various experiments with different shielding compositions are discussed in the reference~\cite{SbyB}. The signal to background ratio depends dominantly on the power of the reactor, overburden and the type of shielding around the detector. 
Due to the segmented geometry of the ISMRAN detector, fast neutron backgrounds can be characterized and discriminated by studying the energy depositions in the PS bars along with the appropriate selection of ratios of the deposited energies in the adjacent PS bars for the prompt IBD event signature. Hence, a detailed knowledge of the fast neutron deposited energy spectrum and response from fast neutrons in the PS bars would enable better optimization of efforts to identify such background events from true IBD events. 
A machine learning approach, using monte-carlo based simulated IBD events in ISMRAN detector, is used to demonstrate the discriminating capabilities of various backgrounds from real IBD events~\cite{MLP}. 
Fast neutron response using $D$-$D$ and $D$-$T$ reactions with plastic scintillator bars are earlier performed to estimate the energy response and sensitivity to the neutron flux in deep underground experiments~\cite{Langford}. 
Fast neutron background arises through cosmogenic neutrons, muon-induced spallation, naturally occurring radioactivity in the local environment, predominantly generated by spontaneous fission of the $\mathrm{{}^{238}U}$ and $\mathrm{{}^{234}Th}$ decay chains from reactor fuel and ($\alpha$,n) reactions in surrounding material~\cite{DBBkgd,Mei,Wulandari,bowden}. Reactor-related fast neutrons typically have energies of a few MeV and neutrons of much higher energies (up to a few GeV) are produced from cosmic neutrons and cosmic muon-induced spallation reactions. The flux and spectrum of fast neutrons from cosmogenic and muon-induced spallations vary with the amount of shielding composition around the detector and by the overburden in the reactor hall~\cite{Cristiano}.

In this paper, we study monoenergetic fast neutron response in plastic scintillator bars using deuterium-deuterium  $D$-$D$ ( 2.45 MeV ) and deuterium-tritium $D$-$T$ ( 14.1 MeV ) reactions at Purnima Neutron Generator Facility (PNGF) at BARC. We present results for energy deposition, timing difference, Z position difference and energy ratio between two adjacent PS bars for the measured data and monte-carlo (MC) generated simulated events. Geant4~\cite{Geant4} package is used to simulate MC events with the physics processes listed in QGSP-BIC-HP and HadronElasticPhysicsHP are used along with the inclusion of photon evaporation model for the de-excitation of Gadolinium nucleus and the resulting $\gamma$-ray cascades. 
Results presented in this work mainly focus on the proton recoil energy deposition and their possible discrimination from the prompt IBD signal consisting of positron energy deposition along with two Compton scattered annihilated $\gamma$-rays in ISMRAN. The event-by-event cascade $\gamma$-rays production from neutron capture on Gadolinium in Geant4 is not properly modeled. However, the total sum energy of the cascade $\gamma$-rays is conserved in the photon evaporation model currently used in this work. Further improvements in this modeling can be achieved by using DICEBOX~\cite{DICEBOX} or by implementing a separate cascade scheme for Gd nucleus from reference~\cite{GdCap} for better description of the neutron capture cascade $\gamma$-rays and is not presented in our current work. The timing, position and energy resolution functions are incorporated in the simulated results when comparing the MC results with data. Simulation studies using IBD and fast neutron events for full ISMRAN detector are used to quantify and reject fast neutron background using the energy ratio in adjacent PS bars. 

\section{{Fast neutron response}:}
Fast neutron undergoes thermalization in organic scintillators. The dominant mechanism of neutron energy loss in scintillator medium is  through elastic scattering from protons, depositing energy in form of proton recoils. The deposited energy from proton recoils in scintillator medium will give light yield output (L) through the scintillation process~\cite{knoll}. In case of $\gamma$-ray, the light yield output response, generated from Compton scattered electrons, is linear over a wide range of energies. That is not the case for heavy charged particles like protons, where the light yield output suffers from quenching effects inside the scintillator medium. For heavy charge particles, which have high specific ionization energy loss, deposit most of their energy in a very small range and the quenching is enhanced by the increased density of $\mathrm{e^{-}}$ and ions. The light yield output response function for heavy charged particles is defined by Birk's formulae :
\begin{equation}\label{eq:birk}
\mathrm{(\frac{dL}{dX}) =  S \times (\frac{dE}{dX}) \times [1 + k_{B} (\frac{dE}{dX}) + C (\frac{dE}{dX})^{2}]^{-1}}, 
\end{equation}
where S is the scintillation efficiency determined through the electron response function, $k_{B}$ and C are adjustable parameters that were empirically determined to fit the light response measurements ~\cite{LO1} for NE-102 polyvinyltoluene plastic scintillator (an equivalent to the EJ-200 used in this work). The values obtained are $\mathrm{k_{B} = 0.011~g~cm^{-2}~MeV^{-1}}$ and C = 9.59 $\times$ $\mathrm{10^{-6} g^{2}~cm^{-4}~MeV^{-2}}$~\cite{Eljen,LO2,LO3}. The specific energy loss used came from the NIST database ~\cite{NIST} for the stopping power and the range of protons in PS bars .


\section{{$\overline\nuup_{e}$ detection through IBD}:}
\begin{figure}
  \centering   
  \includegraphics[width=13cm,height=6cm]{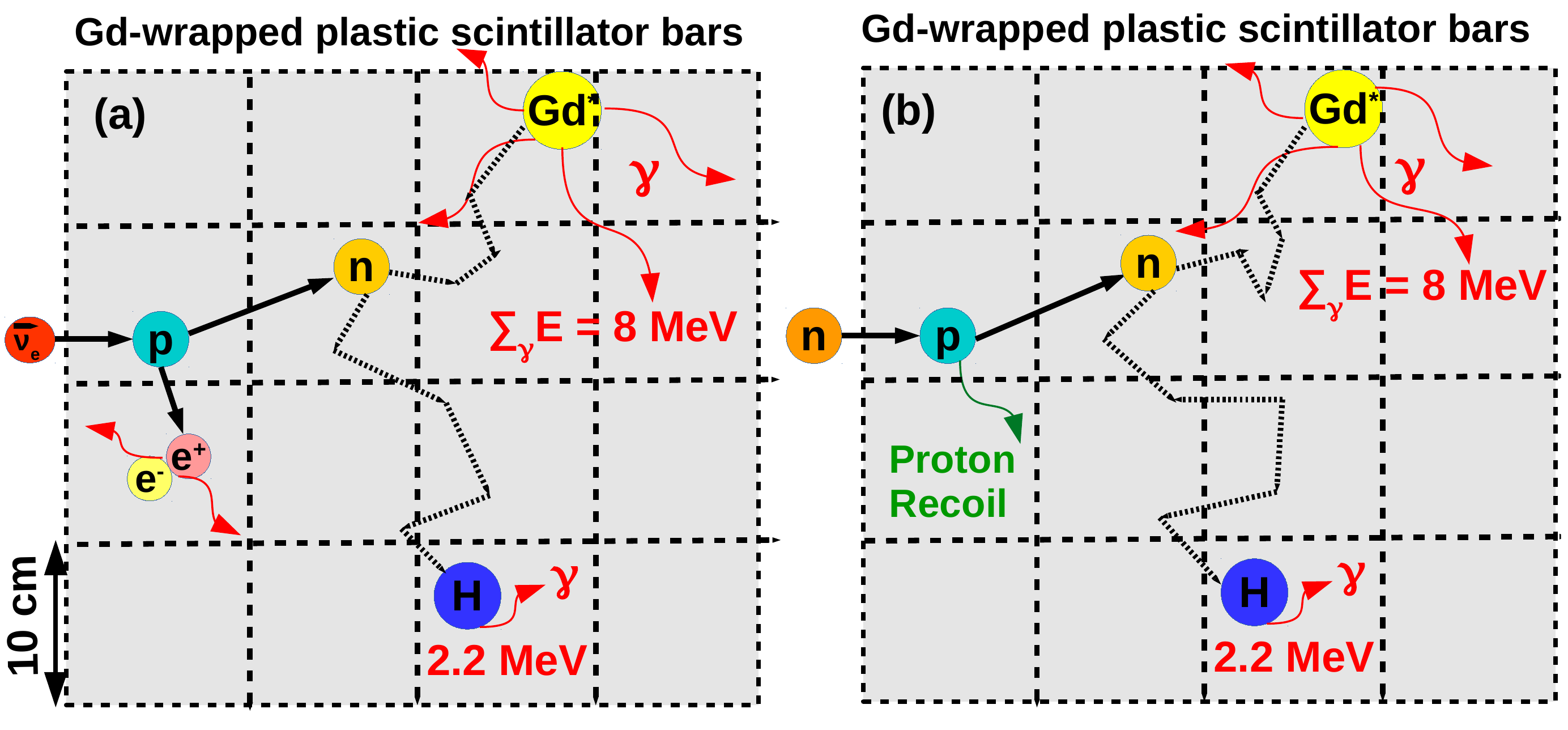}
  \caption{a) Schematic representation of IBD event generating prompt and delayed event signatures in ISMRAN detector. b) Fast neutron event mimicking prompt and delayed event signatures in ISMRAN detector.}
  \label{fig1}
\end{figure}

Reactor $\overline\nuup_{e}$ are primarily detected via the inverse beta decay (IBD) process which has an interaction cross-section of approximately $6 \times 10^{-43} \mathrm{cm^{2}}$. The IBD process is given by,
\begin{equation}\label{eq:ibd}
 \mathrm{ \overline\nuup_{e} + p \rightarrow e^{+} + n}.
\end{equation}
The minimum $\overline\nuup_{e}$ energy required for the above reaction to occur is $\sim$1.8 MeV. In IBD process, positron carries most of the $\overline\nuup_{e}$ energy, and loses its energy in the scintillator volume through ionization. Annihilation of this positron at rest with a surrounding electron produces two $\gamma$-rays of 0.511 MeV. The signal produced by the energy loss of positron and the Compton scattered $\gamma$-rays from the annihilation comprise the prompt event signature. On the other hand, the neutron which has only a few keV's of energy undergoes elastic scattering in the PS bar and thermalize in a timescale of several microseconds ($\mu$s) to get captured on either Gd or H nuclei eq.~(\ref{eq:HC} -~\ref{eq:Gd2}). The thermal neutron capture cross-section on Gd is in several thousands of microbarn and hence the thermal neutron has a greater probability, $\sim$75$\%$, to get captured on Gd in ISMRAN PS bar geometry~\cite{ISMRANNIM}.
\begin{equation}\label{eq:HC}
  \hspace{-0.2in}
\mathrm{ n + p \rightarrow d^{*} \rightarrow \gamma,   \quad E_{\gamma} = 2.2~MeV}, \quad \sigmaup_{n-capture} = \mathrm{0.3~b},
\end{equation}
\begin{equation}\label{eq:Gd1}
  \hspace{-0.2in}
\mathrm{ n + {}^{155}Gd \rightarrow {}^{156}Gd^{*} \rightarrow \gamma 's,   \quad \sum E_{\gamma} = 8.5~MeV}, \quad \sigmaup_{n-capture} = \mathrm{61000~b},
\end{equation}
\begin{equation}\label{eq:Gd2}
\hspace{-0.2in}
\mathrm{ n + {}^{157}Gd \rightarrow {}^{158}Gd^{*} \rightarrow \gamma 's,  \quad \sum E_{\gamma} = 7.9~MeV}, \quad \sigmaup_{n-capture} = \mathrm{254000~b}.
\end{equation}
The de-excitation of Gd nucleus leads to emission of cascade $\gamma$-rays of total energy $\sim$8 MeV. These cascade $\gamma$-rays span more than few PS bars and forms the delayed event signature. The thermal neutron capture on hydrogen in the PS bar produces a monoenergetic $\gamma$-ray of energy 2.2 MeV. The characteristic capture time $\tau$ for Gd-capture and H-capture are $\sim$ $68 \mu$s and $\sim$ $200 \mu$s, respectively.
The time coincidence between prompt and delayed events along with the selection of number of PS bars hit uniquely identifies the IBD event. However, fast neutrons produced from cosmogenic neutrons, muon-induced spallation and from the fission products inside the reactor hall can mimic the prompt event signature by proton recoil energy deposition and the delayed event signature by thermalization inside the ISMRAN detector and get capture on Gd or H nuclei. The thermalization time for these events is very much similar to the IBD events for fast neutrons in energy range of $\sim$15 MeV. Figure~\ref{fig1} shows a schematic representation of IBD (a) and a fast neutron (b) event in ISMRAN detector.

\section{{Experimental setup at PNGF}:}
Deuteron accelerator based neutron generator is developed at BARC~\cite{purnima}. This neutron generator can produce monoenergetic fast neutron with strength of $10^{5}-10^{10}$ neutrons per second which can be operated in both continuous and pulse mode. Neutrons are produced using the following two fusion reactions,
\begin{equation}\label{eq:dd}
  \mathrm{D + D \rightarrow He^{3}(0.82 MeV) + n + 3.3 MeV,~~~~~~E_{n} = 2.45 MeV},
\end{equation}
\begin{equation}\label{eq:dt}
  \mathrm{D + T \rightarrow He^{4}(3.5 MeV)  + n + 17.6 MeV,~~~~~E_{n} = 14.1 MeV}.
\end{equation}
The $D$-$T$ reaction has a larger cross-section, 100 times higher than $D$-$D$ reaction even at low energy from 50-300 keV. This neutron generator (PNGF) has been designed for 300 kV accelerating voltage and 1 mA deuterium beam current installed at Purnima hall. Neutron generator consists of a radiofrequency (RF) ion source. $D^{+}$ ion beam is extracted from RF ion source by applying extraction voltage which is focused by 30 kV DC electrostatic Einzel lens or unipotential lens. At the end of the accelerator,  a target holder assembly is installed which can accommodate Tritium or Deuterium target. Tritium or Deuterium gas adsorption in titanium layer deposited on a copper backing plate of 0.5-1 mm thick is used as a target. Target holder is designed to accommodate 25 to 45 mm diameter target with a cooling system to dissipate beam heat.
\begin{figure}[h!]
  \centering   
  \includegraphics[width=12.5cm,height=6.4cm]{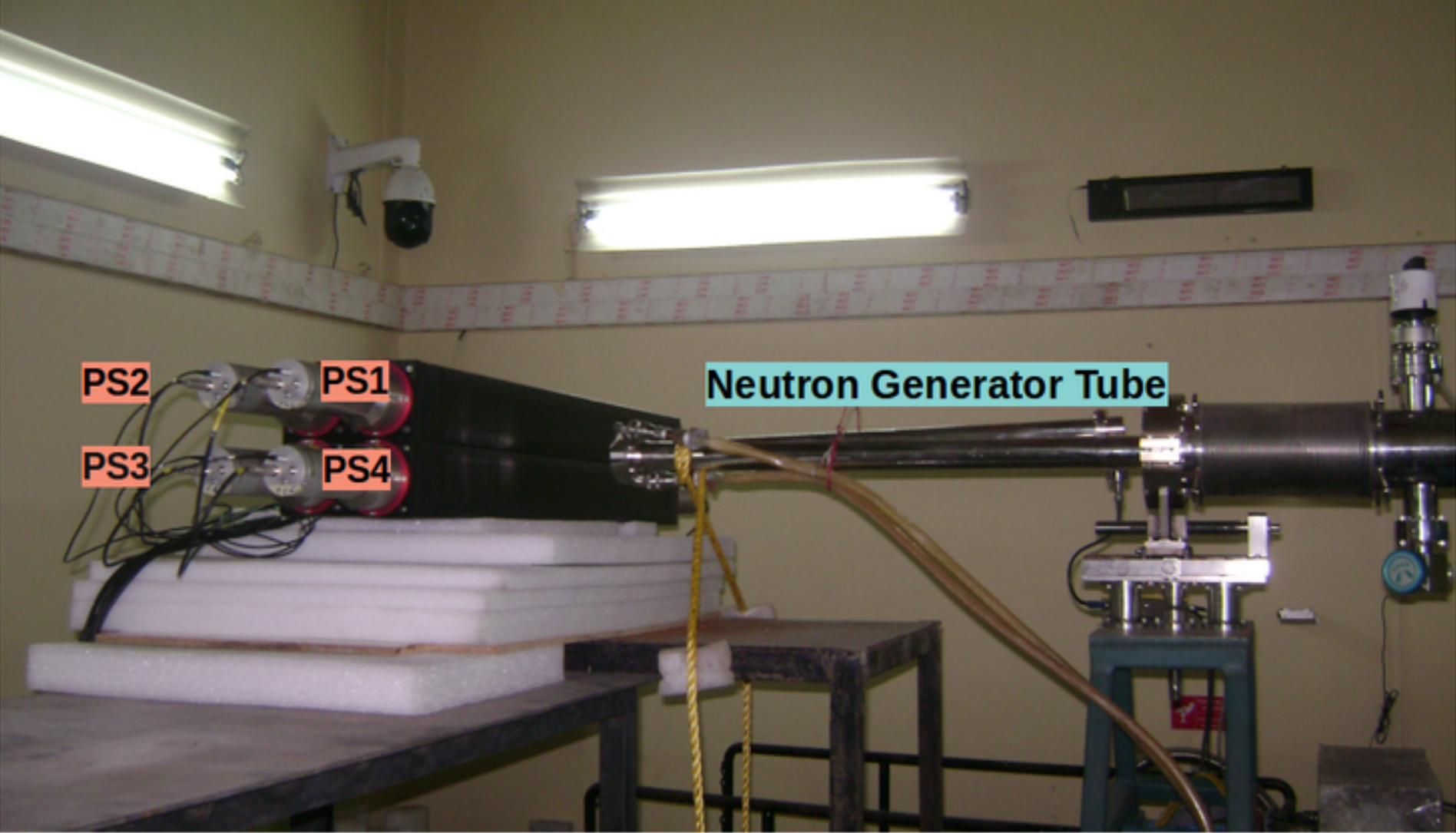}
  \caption{Experimental setup of the PS bars and the 300 kV deuteron accelerator at PNGF.}
  \label{fig2_PNGF}
\end{figure}
The experimental setup, as shown in Fig.~\ref{fig2_PNGF}, consist of 4 PS bars, namely PS1, PS2, PS3, and PS4,  arranged in $2 \times 2$ array. Fast neutron irradiation of the PS bars is performed in PNGF using $D$-$D$ ($E_{n}$ = 2.45 MeV) and $D$-$T$ ($E_{n}$ = 14.1 MeV) fusion reactions. During irradiation, the detectors are placed at 10 cm from the neutron generator beamline and the hit position of the neutron beams are at the center of the PS1 bar. The results in this paper will be discussed from the data recorded from PS1 and its adjacent bar PS2 only as the maximum neutron flux is received by these two bars only. 
Along with PS bars, a 5$''$ NE213 liquid scintillator (LS) detector and a 2$''$ cerium bromide  $\mathrm{CeBr_{3}}$ detector are placed near the plastic scintillator bars to measure neutron and $\gamma$-ray yields in beam ON and OFF conditions, respectively.
A VME based waveform digitizer, CAEN V1730 16 channel 500 MS/sec, is used for the pulse processing and data acquisition system. The pulse discrimination, gate generation, charge integration, timestamp and coincidence of the PMT signals for each PS bar are done on the FPGAs of digitizers. The anode signals from PMTs at both ends of a single bar are required to have a time coincidence of 36 ns to be recorded as a triggered event. The timestamped data from each bar is then further analyzed offline using energy, timing and position information to build events. To obtain uniform response among all the PS bars, a gain matching of PMTs is performed and the energy, timing and position characterization of each PS bar is done independently. The PS bar energy ($\mathrm{E_{bar}}$) is obtained by calibrating each PS bar using $\gamma$-rays and their respective Compton edges from radioactive sources such as $\mathrm{{}^{22}Na, {}^{60}Co, {}^{137}Cs}$ and AmBe neutron source. The timestamp difference ($\mathrm{\Delta T}$) between two PMTs along the length of the bar are used to obtain a parameterization of position (Z) information along each PS bar~\cite{ISMRANNIM}. The middle of each PS bar is considered as the nominal position at 0 cm and both extreme ends are taken as -50cm and +50cm, respectively.
For LS, the pulse shape discrimination (PSD = 1-$Q_{s} / Q_{l})$) parameter is obtained using the charge integrated in short gate of 28 ns ($Q_{s}$) and long gate of 68 ns ($Q_{l}$) to achieve reasonable n-$\gamma$ discrimination. The $\mathrm{CeBr_{3}}$ detector is calibrated using standard radioactive $\gamma$-ray sources and the energy resolution obtained at 0.662 MeV is 3.8$\%$. 
\section{{Results and Discussion}:}
\begin{figure}[h!]
  \hspace{-4.0em}   
  \includegraphics[width=17.4cm,height=5.2cm]{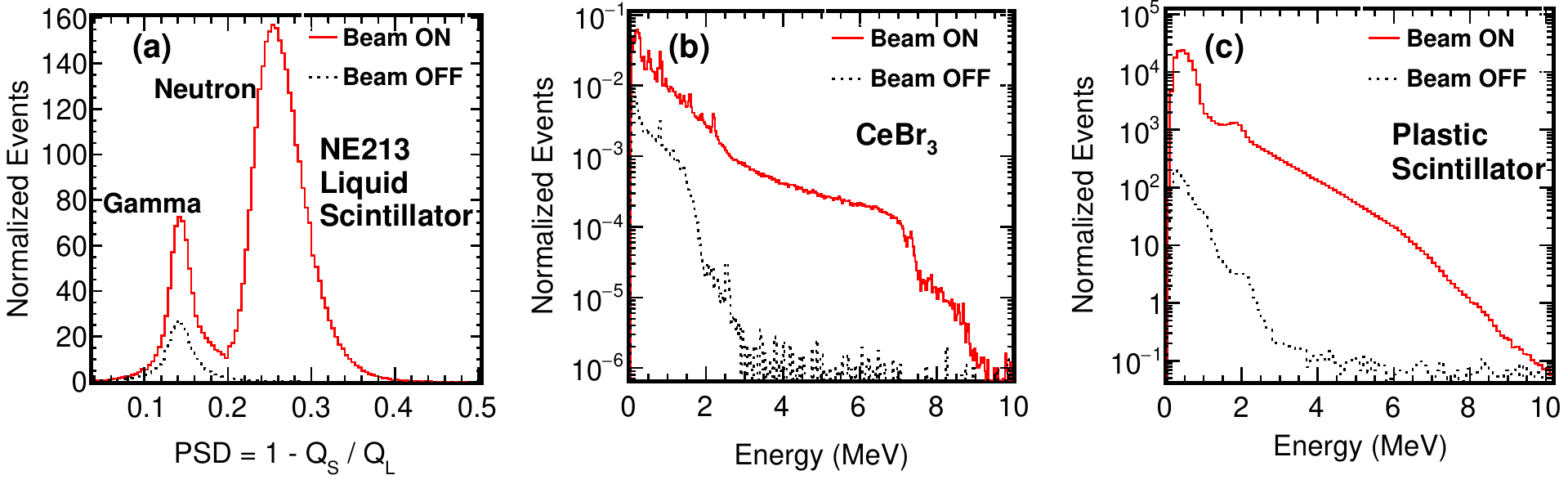}
  \caption{a) PSD distribution for events recorded with NE213 liquid scintillator with beam ON and beam OFF conditions. b) Energy distributions recorded in a $\mathrm{CeBr_{3}}$ $\gamma$-ray detector with beam ON and OFF conditions. c) Energy distributions recorded in a single PS bar with beam ON and OFF conditions for $D$-$D$ reaction.}
  \label{fig3_onoff}
\end{figure}

The $\gamma$-ray and neutron yield during beam ON  and beam OFF conditions for $D$-$D$ reaction at PNGF are studied with $\mathrm{CeBr_{3}}$ and liquid scintillator. Figure~\ref{fig3_onoff} (a) shows the PSD parameter for beam  ON and OFF conditions as measured using liquid scintillator detector. A relatively large neutron yield in front of the PS bars in this configuration is obtained in beam ON condition for $D$-$D$ reaction. On the other hand, a spectral measurement of $\gamma$-ray yield, as shown in Fig.~\ref{fig3_onoff} (b), with $\mathrm{CeBr_{3}}$ compares the relative $\gamma$-ray background in beam ON and OFF conditions. Above 3 MeV in beam ON condition for $D$-$D$ reaction, the energy distribution is dominated by the $\gamma$-rays coming from the neutron capture process on either from Gadolinium wrapped on the PS bars, from iron and copper surrounding support structures and from the shielding material inside the experimental hall ~\cite{Bkg_FN1,Bkg_FN2}. The comparison of the energy spectra shown in Fig~\ref{fig3_onoff} (c), in PS bar for beam ON and OFF conditions in $D$-$D$ reaction shows a proton recoil peak at lower energy and a peak at 2.2 MeV for H capture $\gamma$-ray. 
\begin{figure}[h!]
  \hspace{-1.4em}    
  \includegraphics[width=15.4cm,height=5.973cm]{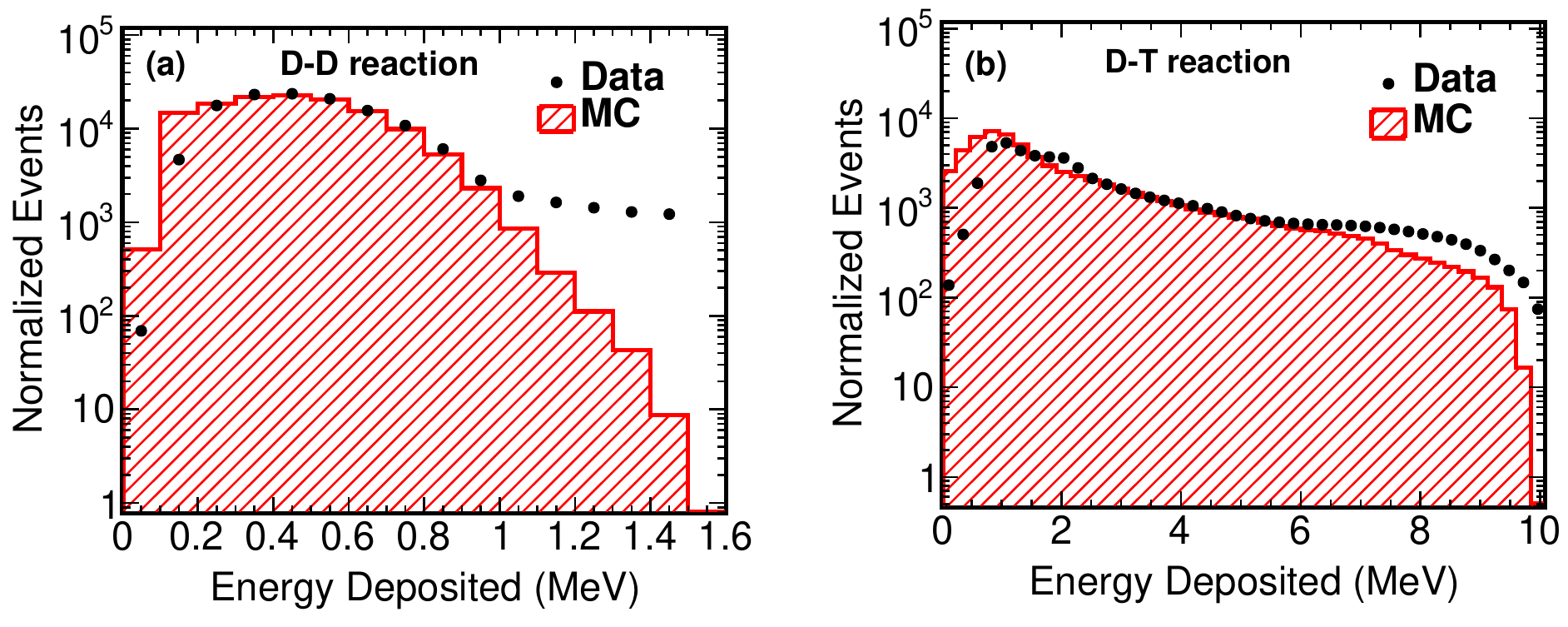}
  \caption{a) Proton recoil energy distribution in PS bar in $D$-$D$ reaction. b) Proton recoil energy distribution in PS bar in $D$-$T$ reaction.}
  \label{fig4_eng_dis}
\end{figure}
Figure~\ref{fig4_eng_dis} (a) and (b) shows the proton recoil energy distribution in PS bar for $D$-$D$ and $D$-$T$ reactions, respectively. The proton recoil energy is expressed in terms of electron equivalent energy $\mathrm{MeV_{ee}}$~\cite{LO4} and now onwards would be referred to as MeV only. The statistical errors on the data are within the symbol size for both $D$-$D$ and $D$-$T$ reactions. The data are compared with the GEANT4 simulated events (MC) to obtain the neutron response at both the energies. The simulation results are smeared with an energy dependent resolution function to get a reasonable agreement with the measured data. At around 2.2 MeV and 7 MeV in $D$-$T$ reaction, MC under predicts data due to the absence of inclusion of neutron capture $\gamma$-rays emanating from either on H, Gd or surrounding material in MC simulations. A selection cut on energy of each PS bar for $D$-$D$ ( 0.1 MeV $< \mathrm{E_{bar}} <$ 1.5 MeV ) and $D$-$T$ ( 0.1 MeV $< \mathrm{E_{bar}} \leq$ 10 MeV ) are imposed to look at the proton recoil events in the adjacent PS bars. Figure~\ref{fig5_deltaT} shows the time correlation ( $\Delta T_{PS2 - PS1}$ ) of events in two bars, PS1 and PS2, in $D$-$D$ and $D$-$T$ reactions. PS1 is the bar which is facing the beam directly and PS2 is the adjacent bar as shown in Fig.~\ref{fig2_PNGF}. This correlation is built by taking the timestamp difference in PS1 and PS2 bars for all the triggered events which satisfy the energy conditions for $D$-$D$ and $D$-$T$ reactions. 
An average of $\sim$12 ns is obtained between the two adjacent PS bars and is mainly due to delays in the signal transfer from the detector to the DAQ system. Also, we have added the same delays in MC simulation of $\Delta T_{PS2 - PS1}$ distributions to be consistent with data. However, the width of $\Delta T_{PS2 - PS1}$, obtained from a Gaussian fit around the peak, is much broader $\sim$3.3 ns in $D$-$D$ reaction as compared to $\sim$1.9 ns in $D$-$T$ reactions. This can be understood in terms of the variation in deposited energy, depending on the incident neutron energy, while traversing through two adjacent PS bars. For $D$-$D$ reactions, due to smaller incident neutron energy, the energy deposition through proton recoil in PS bars is quenched more and hence yields a diminished light output. This introduces larger fluctuations in the collection of photoelectrons at PMTs, due to larger attenuation in scintillation photons energy on PS bars in $D$-$D$ reaction, causing the broader $\Delta T_{PS2 - PS1}$ distribution as compared with $D$-$T$ reaction. From the parameterization of the timestamp differences in each PS bar, the position (Z) of the event is obtained in each individual PS bar~\cite{ISMRANNIM}. Taking the difference ( $\Delta Z_{PS2 - PS1}$ ) of hit positions in adjacent bars, the neutron spread along the beam direction is shown in Fig~\ref{fig6_deltaZ}. The width of $\Delta Z_{PS2 - PS1}$ which gives the neutron coverage in two adjacent bars is 29 cm and 21cm for $D$-$D$ and $D$-$T$ reactions, respectively.
Since the MC events have only the proton recoil component, the disagreement between data and MC events in tail region may be due to the contamination from the $\gamma$-ray background produced from neutron capture on Gd or surrounding material in the beam hall.
\begin{figure}[h!]
  \hspace{-1.4em}    
  \includegraphics[width=15.43cm,height=5.9cm]{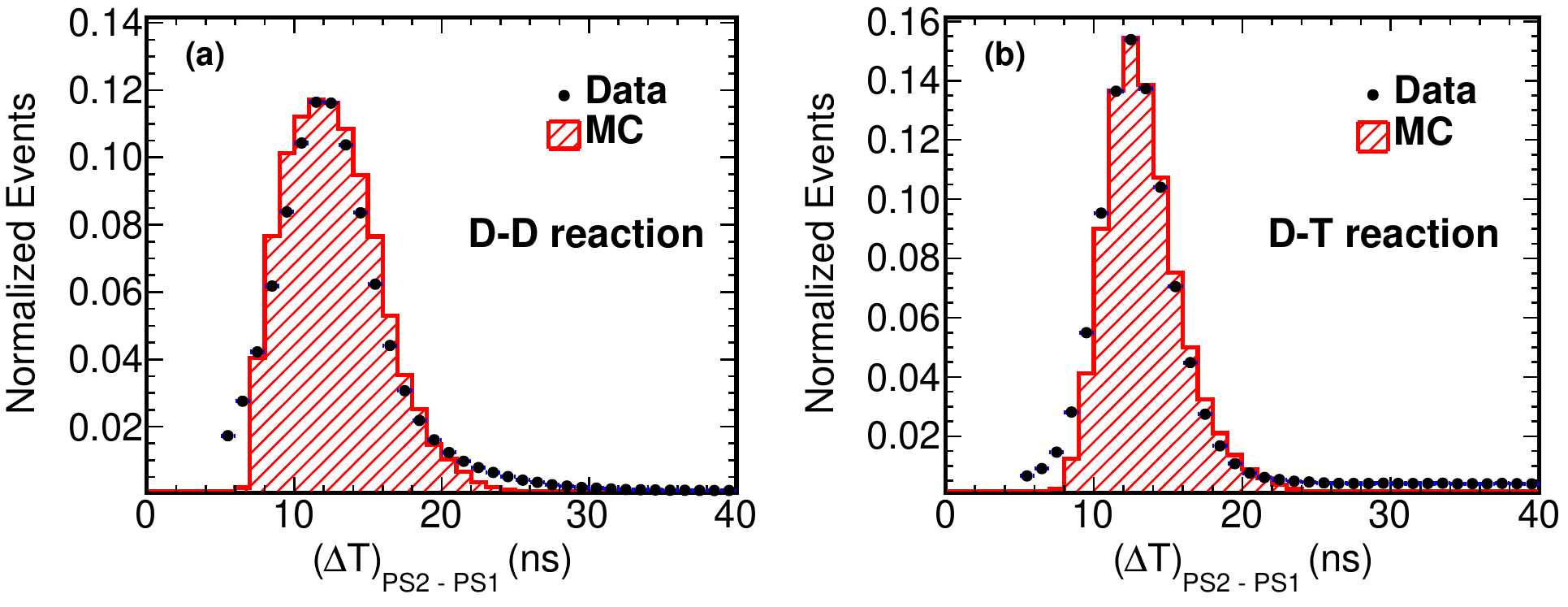}
  \caption{a) $\Delta T_{PS2 - PS1}$ distribution for $D$-$D$ reaction. b) $\Delta T_{PS2 - PS1}$ distribution for $D$-$T$ reaction.}
  \label{fig5_deltaT}
\end{figure}  
\begin{figure}[h!]
  \hspace{-1.1em}   
  \includegraphics[width=15.415cm,height=5.92cm]{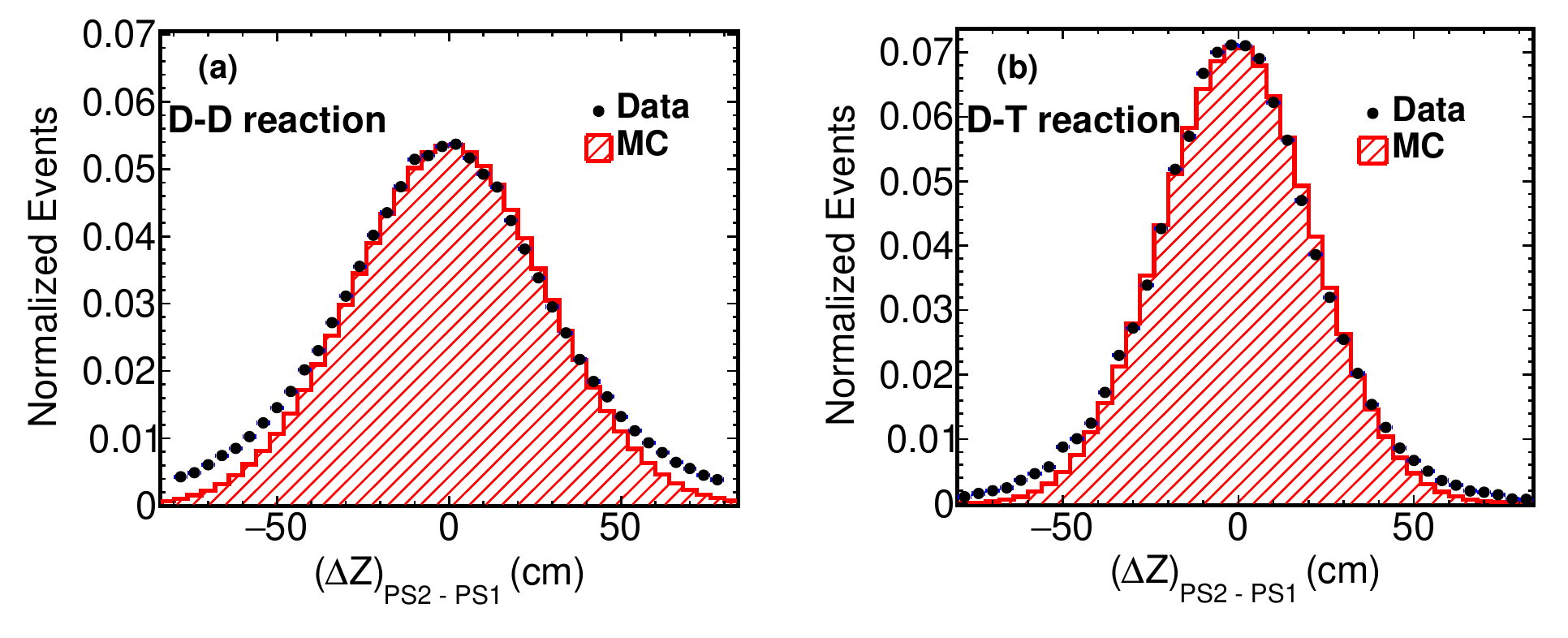}
  \caption{a) $\Delta Z_{PS2 - PS1}$ distribution for $D$-$D$ reaction. b) $\Delta Z_{PS2 - PS1}$ distribution for $D$-$T$ reaction. }
  \label{fig6_deltaZ}
\end{figure}  
The probability, from MC simulations, of a fast neutron thermalizing and undergoing capture on either Gd or H nuclei, which can produce $\gamma$-ray is shown in table~\ref{tab1} for $D$-$D$ and $D$-$T$ reactions. It can be seen from simulations, a higher percentage of events at lower neutron energy would tend to give a fake IBD prompt and delayed event signature as compared to the more energetic fast neutrons. To disentangle the events in adjacent bars from fast neutrons and those from $\gamma$-rays, we study the deposited energy ratio $\mathrm{E_{1}/E_{max}}$ in data and MC simulated events. Since for these studies we are using energy deposition for the two adjacent bars, we use the $\mathrm{E_{max}}$ for the maximum energy deposition and $\mathrm{E_{1}}$ as the next highest energy deposition in either of two bars.
\begin{table}[htbp]
\label{tab1}
\centering
\setlength{\arrayrulewidth}{0.4mm}
\setlength{\tabcolsep}{18pt}

 \begin{tabular}{|c|c|c|c|}
 \hline
 \multicolumn{2}{|l|}{D-D reaction($E_{n} = 2.45 MeV$)}
 &
 \multicolumn{2}{|l|}{D-T reaction($E_{n} = 14.1 MeV$)}\\
 \hline
  H capture & Gd capture   &   H capture & Gd capture \\ [8pt]
  \hline
   0.14$\%$  & 5.10$\%$     & < 0.01$\%$  &   0.82$\%$ \\ [8pt]
   \hline
\end{tabular}
\caption{Thermal neutron capture $\gamma$ events (in $\%$) on H or Gd nuclei giving a signal in two adjacent PS bars in $D$-$D$ and $D$-$T$ reactions from MC simulations.}
\end{table}

\begin{figure}[h!]
  \hspace{-5.0em} 
  \includegraphics[width=17.8cm,height=5.8cm]{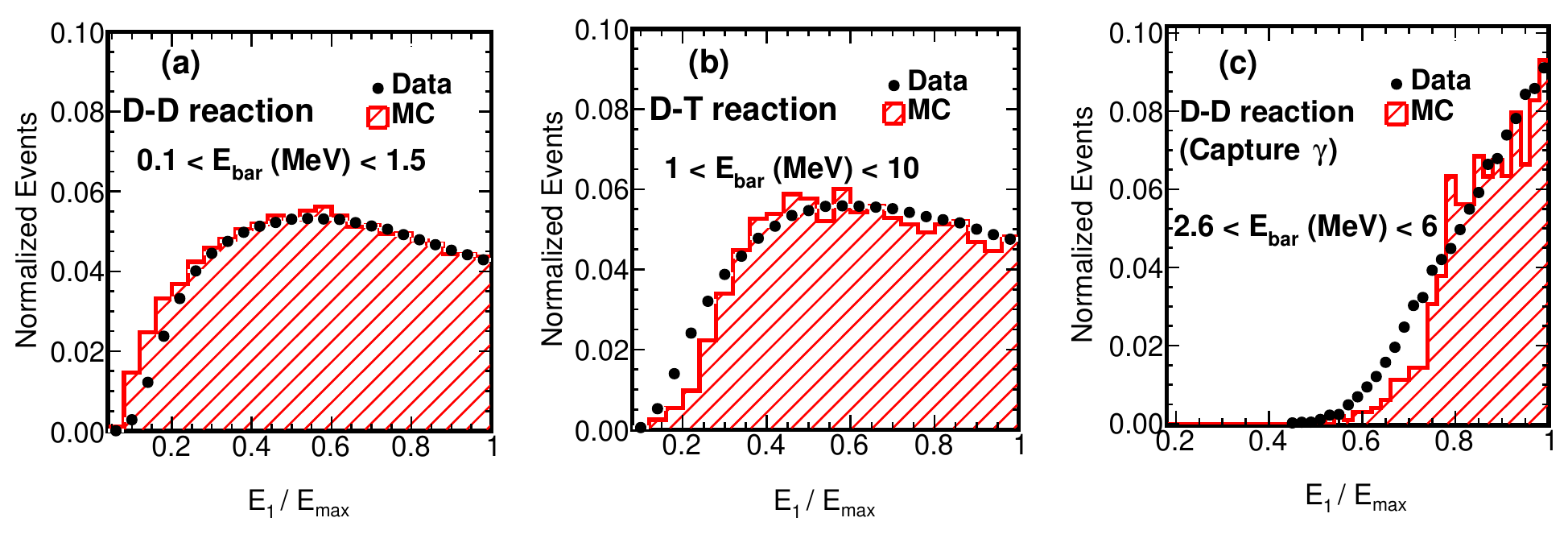}
  \caption{Proton recoil energy ratio distribution between PS1 and PS2 for (a) $D$-$D$ and (b) $D$-$T$ reactions between PS1 and PS2 bars. c) Energy ratio distribution between PS1 and PS2 for $\gamma$-ray background in $D$-$D$ reaction with energy selection of 2.6 $< \mathrm{E_{bar}} < $ 6.0 MeV in each PS bar.}
  \label{fig7_eratio}
\end{figure}  
Figure~\ref{fig7_eratio} (a) and (b) shows the distribution of ratio of energy deposition ($\mathrm{E_{1}/E_{max}}$) between PS2 and PS1 for $D$-$D$ and $D$-$T$ reactions with an energy selection of 0.1 MeV $< \mathrm{E_{bar}} <$ 1.5 MeV and 1 MeV $< \mathrm{E_{bar}} <$ 10 MeV on each bar, respectively.  The energy ratio $\mathrm{(E_{1}/E_{max})}$ results for data and MC simulations agree well in the range of $\mathrm{E_{bar}}$ for  $D$-$D$ and $D$-$T$ reactions. In $D$-$D$ reactions, above 2.45 MeV, the energy deposited in the adjacent bars is purely due to the $\gamma$-rays from neutron capture on Gd or surrounding material. To study the characteristic features of energy ratio $\mathrm{(E_{1}/E_{max})}$ within the adjacent PS bars exclusively for $\gamma$-ray background, we select events with energy  2.6 MeV $< \mathrm{E_{bar}} <$ 6.0 MeV for $D$-$D$ reactions. Figure~\ref{fig7_eratio} (c) shows the comparison of energy ratio $\mathrm{(E_{1}/E_{max})}$ between data and MC events within the energy range 2.6 MeV $< \mathrm{E_{bar}} <$ 6.0 MeV for $D$-$D$ reactions. There seems to be a disagreement between data and MC in the region from 0.4 to 0.8 which may be due to the poor modeling of the n-Gd cascade $\gamma$-rays in Geant4 PhotonEvaporation model. However, the shape of energy ratio $\mathrm{(E_{1}/E_{max})}$ between the fast neutron events and $\gamma$-ray background events are different and hence can be used for discriminating the fast neutron events from $\gamma$-ray events by making an appropriate selection of $\mathrm{E_{1}/E_{max}}$.
\begin{figure}[h!]
   \hspace{-4.0em} 
  \includegraphics[width=17.4cm,height=5.2cm]{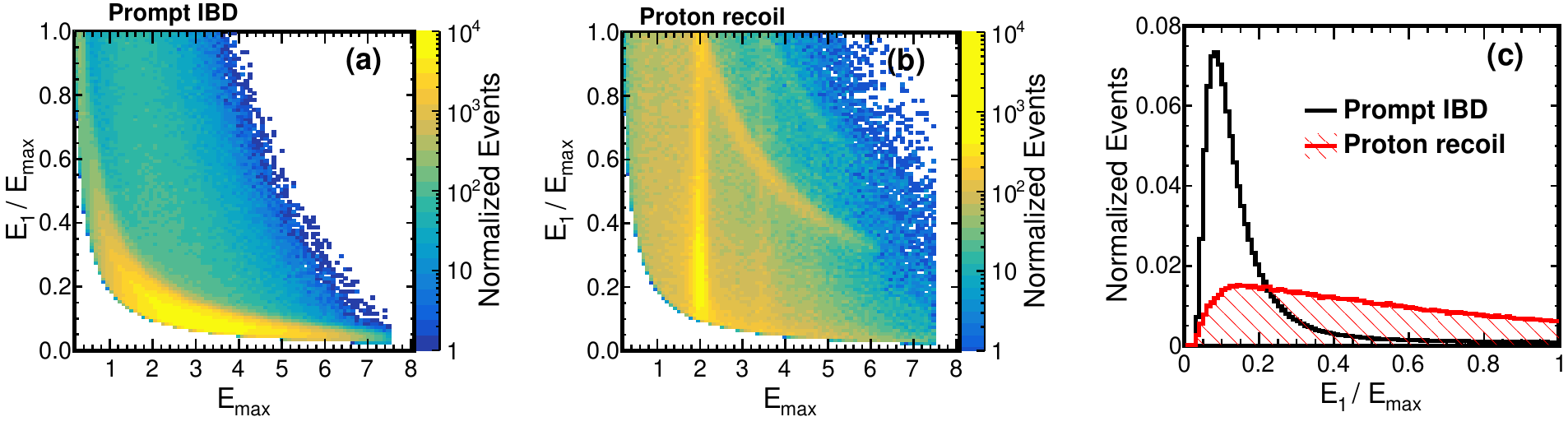}
  \caption{ Energy ratio $\mathrm{(E_{1} / E_{max})}$ vs. maximum energy deposition $\mathrm{ E_{max} }$ between two adjacent PS bars for prompt event from ~a) positron ionization from IBD, ~ b) proton recoil from fast neutron.  ~c) Projected distribution of deposited energy ratio $\mathrm{(E_{1} / E_{max})}$ between two adjacent PS bars for prompt positron event from IBD and proton recoil event from fast neutrons. }
  \label{fig8_crop}
\end{figure}

Using above obtained results from fast neutron in $D$-$D$ and $D$-$T$ reactions, we study the energy ratio $\mathrm{(E_{1}/E_{max})}$ of two adjacent bars for true IBD prompt events and fast neutron events in full ISMRAN detector using the MC simulated events. The fast neutrons are generated with a uniform energy distribution from 2 to 12 MeV randomizing the position of the event vertex in the ISMRAN detector. This assumption is taken from RENO experiment~\cite{FN_RENO} which is a good approximation of the fast neutron yield in the studied energy range. From MC simulations, a weak dependence is observed between the energy ratio $\mathrm{(E_{1}/E_{max})}$ and the incident neutron energy within the studied energy range of fast neutrons. More accurate description of the fast neutron energy dependence will be available once the measurements of fast neutron will be made inside the 10 cm of Pb and 10 cm of BP shielding structure at the experimental site in the reactor hall. 
The IBD events are generated randomly in ISMRAN detector, using anti-neutrino energy dependent parameterization obtained from the following references~\cite{Huber,Mueller}. The prompt positron and neutron are produced kinematically according to the $\overline\nuup_{e}$ energy. The energy loss, propagation and energy depositions in the ISMRAN volume are recorded using the Geant4 standard packages mentioned earlier in this paper.
The proton recoil from fast neutrons can mimic like a prompt IBD event signature in the ISMRAN detector in terms of total energy deposition and number of bars hit. The energy ratio $\mathrm{(E_{1} / E_{max})}$ may be used as a discriminating variable for separation of the signal prompt IBD events from background fast neutron events, since the energy profiles of both events are different in nature with respect to the segmentation of the detector. On the other hand, the energy profile and energy ratio $\mathrm{(E_{1} / E_{max})}$ for delayed event signature would be identical between the delayed capture IBD event and fast neutron thermalized captured event ( can be seen from Fig~\ref{fig1} ) and would not be useful for differentiating the fast neutron background events from IBD events. For this study, we have only considered those PS bars where the energy deposition is between 0.2 to 7.5 MeV. This energy selection denotes the region of interest for the measurement of prompt and delayed IBD event signatures above the experimental threshold of PS bar and in the reactor $\overline\nuup_{e}$ energy range.     
Figure~\ref{fig8_crop} (a) and (b) shows the correlation of energy ratio $\mathrm{(E_{1} / E_{max})}$ as a function of $\mathrm{E_{max}}$ for prompt IBD events and the proton recoil event signature from fast neutron events, respectively. For prompt IBD event, the energy ratio $\mathrm{(E_{1} / E_{max})}$ has a smooth profile as a function of $\mathrm{E_{max}}$ as opposed to that for the proton recoil fast neutron events. Subsequent bands at 2 MeV and 3.5 MeV in the proton recoil events are due to the energy thresholds applied in the selection of $\mathrm{E_{1}}$ and $\mathrm{E_{max}}$. We have explicitly confirmed this in MC simulations by lowering the energy threshold to 0.01 on individual bars.  
Figure~\ref{fig8_crop} (c) shows the projected distribution of deposited energy ratio $\mathrm{(E_{1} / E_{max})}$ between two adjacent PS bars for prompt positron event from IBD and proton recoil event from fast neutrons. Since most of the anti-neutrino energy is carried by positron and due to the rapid ionization energy loss of the positron in PS bar, this energy deposition is mostly restricted to single PS bar. On the other hand, the annihilation $\gamma$-rays, with maximum of 1.022 MeV, can span multiple bars giving a unique feature in the energy ratio $\mathrm{(E_{1} / E_{max})}$ for prompt IBD events. For fast neutrons, the proton recoil energy deposition is random and we observe a broad $\mathrm{(E_{1} / E_{max})}$ distribution. By selecting events based on the energy ratio $\mathrm{(E_{1} / E_{max})}<$ 0.3, yields a selection efficiency of 88$\%$ for the prompt IBD events with a rejection of 64$\%$ of the fast neutron events from MC simulation studies. Stricter selection criteria based on event topology in segmented geometry and timing distribution among the PS bars can further improve the efficiency of prompt IBD events while rejecting the fast neutron background events. 
A machine learning approach using multi-layer perceptrons, the fast neutron background is discriminated with an efficiency of 80$\%$ from the prompt IBD signal~\cite{MLP}. However, due to small number of bars used in the current experimental setup, results from such a approach would be difficult to interpret and hence are not discussed in present work.

\section{{Conclusions and Outlook}:}
The fast neutron response, with an array ($2 \times 2$) of plastic scintillator bars, is studied using monoenergetic fast neutron from $D$-$D$ and $D$-$T$ reactions. The response of energy deposition from proton recoil events are presented for the plastic scintillator bars. A reasonable agreement between data and monte-carlo simulated events for the energy deposition in PS bars from proton recoil events are observed. The segmented geometry using PS bars gives an advantage in discriminating the proton recoil events from the $\gamma$-ray background events by studying the ratio of the deposited energy in adjacent PS bars. A validation between data and monte-carlo events is obtained for the energy ratio $\mathrm{(E_{1} / E_{max})}$ for proton recoil events. The energy ratio $\mathrm{(E_{1} / E_{max})}$ from fast neutron background events and $\gamma$-ray background events can be discriminated against IBD prompt events using the energy depositions in adjacent bars in ISMRAN detector. For the full ISMRAN geometry, using monte-carlo simulated events, it is observed that an efficiency of 88$\%$ for prompt IBD events can be obtained with a rejection of 64$\%$ of fast neutron events by selecting events $\mathrm{(E_{1} / E_{max})} < $ 0.3. Further improvement in the separating the prompt IBD events can be achieved by combining energy ratio $\mathrm{(E_{1} / E_{max})}$ variable with other topological event selection cuts in PS bars along with the implementation of an advanced machine learning algorithms.

\section{{Acknowledgements}:}
We are thankful to Neutron $\&$ X-ray Physics Division (NXPD), BARC for logistical support and co-operation during data taking.


\begin{thebibliography}{99}
\bibitem{DBMIX}
  F.~P.~An et~al. (Daya Bay Collaboration), \emph{Spectral Measurement of Electron Antineutrino Oscillation Amplitude and Frequency at Daya Bay}, \emph{Phys. Rev. Lett.}, {\bf 112} (2014) 061801.\\
  \url{https://doi.org/10.1103/PhysRevLett.112.061801}.
  
\bibitem{RENO}
  J.~H.~Choi et~al. (RENO Collaboration), \emph{Observation of Energy and Baseline Dependent Reactor Antineutrino Disappearance in the RENO Experiment}, \emph{Phys. Rev. Lett.}, {\bf 116} (2016) 211801. 
  \url{https://doi.org/10.1103/PhysRevLett.116.211801}.
  
\bibitem{DChooz}
  Abe.~Y. et~al. (Double Chooz Collaboration), \emph{Indication of Reactor ${\overline{\ensuremath{\nu}}}_{e}$ Disappearance in the Double Chooz Experiment}, \emph{Phys. Rev. Lett.}, {\bf 108} (2012) 131801.
  \url{https://link.aps.org/doi/10.1103/PhysRevLett.108.131801}.     
  

\bibitem{DANSS}
  I.~Alekseev. et~al. (DANSS Collaboration), \emph{Detector of the reactor AntiNeutrino based on Solid Scintillator}, \emph{Journal of Instrumentation}, {\bf 11} (2016) P11011.
  \url{http://stacks.iop.org/1748-0221/11/i=11/a=P11011}. 

\bibitem{NEOS}
  Y.~J.~Ko et~al. (NEOS Collaboration), \emph{Sterile Neutrino Search at the NEOS Experiment}, \emph{Phys. Rev. Lett.}, {\bf 118} (2017) 121802.
  \url{https://doi.org/10.1103/PhysRevLett.118.121802}.
  
\bibitem{RENOST}
  J.~H.~Choi  et~al. (RENO Collaboration), \emph{Search for Sub-eV Sterile Neutrinos at RENO}, \emph{Phys. Rev. Lett.}, {\bf 125} (2020) 191801.
  \url{https://journals.aps.org/prl/abstract/10.1103/PhysRevLett.125.191801}.  
  

\bibitem{STEREO}
  H.~Almazan et~al. (STEREO Collaboration), \emph{Accurate Measurement of the Electron Antineutrino Yield of $\mathrm{{}^{235}U}$ Fissions from the STEREO Experiment with 119 Days of Reactor-On Data}, \emph{Phys. Rev. Lett.}, {\bf 125} (2020) 201801.
  \url{https://journals.aps.org/prl/abstract/10.1103/PhysRevLett.125.201801}. 
  
\bibitem{PROSPECT}
  J. Ashenfelter. et~al. (The PROSPECT Collaboration), \emph{Measurement of the Antineutrino Spectrum from $\mathrm{{}^{235}U}$ Fission at HFIR with PROSPECT}, \emph{Phys. Rev. Lett.}, {\bf 122} (2019) 251801.
  \url{https://journals.aps.org/prl/abstract/10.1103/PhysRevLett.122.251801}.

\bibitem{SOLID}
  Y. Abreu. et~al. (The SoLid Collaboration), \emph{Performance of a full scale prototype detector at the BR2 reactor for the SoLid experiment,}, \emph{Journal of Instrumentation}, {\bf 13} (2018) P05005.
  \url{https://iopscience.iop.org/article/10.1088/1748-0221/13/05/P05005}. 
  
\bibitem{RENO5MeV}
  G.~Bak et~al. (RENO Collaboration), \emph{Fuel-Composition Dependent Reactor Antineutrino Yield at RENO}, \emph{Phys. Rev. Lett.}, {\bf 122} (2019) 232501.
  \url{https://journals.aps.org/prl/abstract/10.1103/PhysRevLett.122.232501}.
  
\bibitem{DB5MeV}
  F.~P.~An et~al. (Daya Bay Collaboration), \emph{Evolution of the Reactor Antineutrino Flux and Spectrum at Daya Bay}, \emph{Phys. Rev. Lett.}, {\bf 118} (2017) 251801.
  \url{https://journals.aps.org/prl/abstract/10.1103/PhysRevLett.118.251801}. 
  
\bibitem{Huber5MeV}
  Patrick.~ Huber et~al. \emph{NEOS Data and the Origin of the 5 MeV Bump in the Reactor Antineutrino Spectrum}, \emph{Phys. Rev. Lett.}, {\bf 118} (2017) 042502.
  \url{https://journals.aps.org/prl/abstract/10.1103/PhysRevLett.118.042502}.   

\bibitem{DHRUVA}
  S.~K.~Agarwal et~al. \emph{Dhruva: Main design features, operational experience and utilization}, \emph{Nuclear Engineering and Design}, {\bf 236} (2006) 747-757.\url{https://www.sciencedirect.com/science/article/abs/pii/S0029549306000732?via%3Dihub}.    
  
\bibitem{ISMRANNIM}
  D.~Mulmule. et~al. (ISMRAN Collaboration), \emph{A plastic scintillator array for reactor based anti-neutrino studies}, \emph{Nuclear Instruments and Methods in Physics Research Section A: Accelerators, Spectrometers, Detectors and Associated Equipment}, {\bf 911} (2018) 104-114.
  \url{https://www.sciencedirect.com/science/article/abs/pii/S0168900218313408#!}.

\bibitem{IAEA}
Technical Meeting on Nuclear Data for Anti-neutrino Spectra and Their Applications, 23-26 April 2019, IAEA Headquarters, Vienna, Austria.
 \url{https://www-nds.iaea.org/index-meeting-crp/Antineutrinos/}.

\bibitem{PANDA}
  S.~Oguri. et~al. (PANDA Collaboration), \emph{Reactor antineutrino monitoring with a plastic scintillator array as a new safeguards method}, \emph{Nuclear Instruments and Methods in Physics Research Section A: Accelerators, Spectrometers, Detectors and Associated Equipment}, {\bf 757} (2014) 33-39.
  \url{https://www.sciencedirect.com/science/article/abs/pii/S0168900214004781}.  

\bibitem{Shiba}
  S.~P.~Behera et~al. (ISMRAN Collaboration), \emph{Active-sterile neutrino mixing constraints using reactor antineutrinos with the ISMRAN setup}, \emph{Phys. Rev. D}, {\bf 102} (2020) 013002.
  \url{https://journals.aps.org/prd/abstract/10.1103/PhysRevD.102.013002}.
 
 \bibitem{Reactor bkg}
J.~Ashenfelter et. al. (The PROSPECT Collaboration) \emph{Background Radiation Measurements at High Power Research Reactors}, \emph{Nuclear Instruments and Methods in Physics Research Section A: Accelerators, Spectrometers, Detectors and Associated Equipment}, {\bf 806} (2016) 401-419.
\url{https://www.sciencedirect.com/science/article/abs/pii/S0168900215012309}

\bibitem{DBBkgd}
  F.~P.~AN et~al. (The Daya Bay Collaboration), \emph{Cosmogenic neutron production at Daya Bay}, \emph{Phys. Rev. D}, {\bf 97} (2018) 052009.
  \url{https://journals.aps.org/prd/abstract/10.1103/PhysRevD.97.052009}. 

\bibitem{Mei}
  D.~ M.~ Mei et~al. \emph{Muon-induced background study for underground laboratories}, \emph{Phys. Rev. D}, {\bf 73} (2006) 053004.
  \url{https://journals.aps.org/prd/abstract/10.1103/PhysRevD.73.053004}.

\bibitem{Wulandari}
 H. ~Wulandari et~al. \emph{Neutron Background Studies for the CRESST Dark Matter Experiment}, ArXiv High Energy Physics Experiment e-prints arXiv:hep-ex/0401032.
 \url{https://arxiv.org/pdf/hep-ex/0401032.pdf}

\bibitem{bowden}
  N.S.~Bowden. et~al. \emph{A note on neutron capture correlation signals, backgrounds, and efficiencies}, \emph{Nuclear Instruments and Methods in Physics Research Section A: Accelerators, Spectrometers, Detectors and Associated Equipment}, {\bf 693} (2012) 209-214.
  \url{https://www.sciencedirect.com/science/article/abs/pii/S0168900212007528}.
   
\bibitem{SbyB}
 Yeongduk.~Kim~ et~al. \emph{Detection of Antineutrinos for Reactor Monitoring}, \emph{Nuclear Engineering and Technology}, {\bf 48} (2016) 285-292.
\url{https://www.sciencedirect.com/science/article/pii/S1738573316000498}.


\bibitem{MLP}
  D.~Mulmule. et~al. (ISMRAN Collaboration), \emph{Machine learning technique to improve anti-neutrino detection efficiency for the ISMRAN experiment}, \emph{Journal of Instrumentation}, {\bf 15} (2020) P04021.
  \url{https://iopscience.iop.org/article/10.1088/1748-0221/15/04/P04021}. 


 \bibitem{Langford}
  T.~J.~Langford et~al. \emph{Fast neutron detection with a segmented spectrometer}, \emph{Nucl. Instrum. Meth. A}, {\bf 771} (2015) 78-87.
 [arXiv:1407.6601]
\url{https://www.sciencedirect.com/science/article/abs/pii/S0168900214012170#!}.
  

\bibitem{Cristiano}
  Cristiano.~ Galbiati et~al. \emph{Measuring the cosmic ray muon-induced fast neutron spectrum by (n,p) isotope production reactions in underground detectors}, \emph{Phys. Rev. C}, {\bf 72} (2005) 025807[Erratum ibid.
73 (2006) 049906].
  \url{https://journals.aps.org/prc/abstract/10.1103/PhysRevC.72.025807}.

  
\bibitem{Geant4}
  S.~Agostinelli et~al. \emph{Geant4 - a simulation toolkit}, \emph{Nuclear Instruments and Methods in Physics Research Section A: Accelerators, Spectrometers, Detectors and Associated Equipment}, {\bf 506} (2003) 250-303.
  \url{http://www.sciencedirect.com/science/article/pii/S0168900203013688}.   

\bibitem{DICEBOX}
F.~Becvar, \emph{Simulation of $\gamma$ cascades in complex nuclei with emphasis on assessment of uncertainties of cascade-related quantities}, \emph{Nuclear Instruments and Methods in Physics Research Section A}, {\bf 417} (1998) 434-449. 
\url{https://www.sciencedirect.com/science/article/abs/pii/S0168900298007876}.

\bibitem{GdCap}
Kaito.~ Hagiwara  et~al. \emph{Gamma-ray spectrum from thermal neutron capture on gadolinium-157}, \emph{Progress of Theoretical and Experimental Physics}, {\bf 2019} (2019) 023D01.
\url{https://academic.oup.com/ptep/article/2019/2/023D01/5362643?login=true}.

\bibitem{knoll}
  G. Knoll. et~al. \emph{Radiation Detection and Measurement}, \emph{John Wiley and Sons, Inc.}, (2000).
  \url{https://www.fulviofrisone.com/attachments/article/444/Radiation%20Detection%20and%20Measurement,%203rd%20ed%20-%20Glenn%20F.%20Knoll%20(Wiley,%202000).pdf}.
 
\bibitem{LO1}
  V.~I.~Tretyak. ~ et~al. \emph{Semi-empirical calculation of quenching factors for ions in scintillators}, \emph{Nuclear Instruments And Methods }, {\bf 33} (2010) 40-53.
  \url{https://www.sciencedirect.com/science/article/abs/pii/S0927650509001650}.
 
\bibitem{Eljen}
  Eljen Technology, https://eljentechnology.com/products/plastic-scintillators.

\bibitem{LO2}
  Richard.~ Madey. ~ et~al. \emph{The response of NE-228A, NE-228, NE-224, and NE-102 scintillators to protons from 2.43 to 19.55 MeV}, \emph{Nuclear Instruments And Methods }, {\bf 151} (1978) 445-450.
  \url{https://www.sciencedirect.com/science/article/abs/pii/0029554X78901544#!}.

\bibitem{LO3}
  R.A. Cecil, B.D. Anderson and R. Madey \emph{Improved predictions of neutron detection efficiency for hydrocarbon scintillators from 1 MeV to about 300 MeV}, \emph{Nuclear Instruments And Methods }, {\bf 161} (1979) 439-447.
  \url{https://www.sciencedirect.com/science/article/abs/pii/0029554X79904178}.
  
\bibitem{NIST}
NIST Database, \url{https://www.nist.gov/pml/stopping-power-range-tables-electrons-protons-and-helium-ions}. 

\bibitem{purnima}
  T. Patel. et~al. \emph{D-D/D-T neutron generator facilities for basic and applied research}, (2015).  
  \url{http://sympnp.org/proceedings/63/G46.pdf}.

\bibitem{Bkg_FN1}
  C.~ M.~ Eisenhauer ~ et~al. \emph{Review of Scattering Corrections for Calibration of Neutron Instruments}, \emph{Radiation Protection Dosimetry}, {\bf 28} (1989) 253-262.
  \url{https://academic.oup.com/rpd/article-abstract/28/4/253/1604986?redirectedFrom=fulltext}.

\bibitem{Bkg_FN2}
  Sang.~ In.~ Kim ~ et~al. \emph{A review of neutron scattering correction for the calibration of neutron survey meters using the shadow cone method}, \emph{Nuclear Engineering and Technology}, {\bf 47} (2015) 939-944.
  \url{https://www.sciencedirect.com/science/article/pii/S173857331500176X#!}.

\bibitem{LO4}
  J.~ A.~ Brown ~ et~al. \emph{Proton Light Yield in Organic Scintillators using a Double Time-of-Flight Technique}, {\bf 124} (2018) 045101.
  \url{https://doi.org/10.1063/1.5039632}.

\bibitem{FN_RENO}
 S.~H.~Seo. ~ et~al. \emph{Spectral measurement of the electron antineutrino oscillation amplitude and frequency using 500 live days of RENO data}, \emph{Phys. Rev. D}, {\bf 98} (2018) 012002.
\url{https://journals.aps.org/prd/abstract/10.1103/PhysRevD.98.012002}.

\bibitem{Huber}
  Patrick.~Huber~ et~al. \emph{Determination of antineutrino spectra from nuclear reactors}, \emph{Phys. Rev. C}, {\bf 84} (2011) 024617.
  \url{https://journals.aps.org/prc/abstract/10.1103/PhysRevC.84.024617}.
  
\bibitem{Mueller}
  Th.~ A.~ Mueller et~al. \emph{Improved predictions of reactor antineutrino spectra}, \emph{Phys. Rev. C}, {\bf 83} (2011) 054615.
  \url{https://journals.aps.org/prc/abstract/10.1103/PhysRevC.83.054615}.
  










     

\end{thebibliography}
\end{document}